\RequirePackage{lineno}
\documentclass[twocolumn,showpacs,superscriptaddress,amsmath,amssymb,nofootinbib]{revtex4-2}

\usepackage{graphicx}
\usepackage{subfigure}
\usepackage{subcaption}
\usepackage{epsfig}
\usepackage{overpic}
\usepackage{dcolumn}
\usepackage{ulem}
\usepackage{bm}
\usepackage{color}
\usepackage{lineno}
\usepackage{xspace}
\usepackage{multirow}
\usepackage{epstopdf}
\usepackage{xcolor}
\usepackage{soul}
\usepackage{verbatim}
\usepackage{enumitem}
\usepackage{todonotes}
\usepackage{cancel}
\usepackage{array}
\usepackage[toc,page]{appendix}
\usepackage[colorlinks,allcolors=black]{hyperref}

\usepackage{diagbox}

\newcommand{\OU}{\affiliation{Research Center for Nuclear Physics (RCNP), Osaka University, Ibaraki 567-0047, Japan}}

  \newcommand{\moe}{\affiliation{Key Laboratory of Atomic and Subatomic Structure and Quantum Control (MOE), Guangdong-Hong Kong Joint Laboratory of Quantum Matter, Guangzhou 510006, China
}}

\newcommand{\sfim}{\affiliation{Guangdong Basic Research Center of Excellence for Structure and Fundamental Interactions of Matter, Guangdong Provincial Key Laboratory of Nuclear Science, Guangzhou 510006, China}}

\newcommand{\iqm}{\affiliation{State Key Laboratory of Nuclear Physics and Technology, Institute of Quantum Matter, South China Normal University, Guangzhou 510006, China}}

\newcommand{\scnt}{\affiliation{Southern Center for Nuclear-Science Theory (SCNT), Institute of Modern Physics, Chinese Academy of Sciences, Huizhou 516000, Guangdong Province, China}}

\newcommand{\shanghai}{\affiliation{College of Science, University of Shanghai for Science and Technology, Shanghai 200093, China}}

\newcommand{\gscas}{\affiliation{Graduate School of China Academy of Engineering Physics, Beijing 100193, China}}

\begin{document}
\include{def-com}
\title{\boldmath Machine Learning Unveils the Power Law of Finite-Volume Energy Shifts}

\author {Wei-Jie Zhang}
\iqm
\moe
\sfim
\author {Zhenyu Zhang} \email{Co-first author}
\iqm
\moe
\sfim
\author {Jifeng Hu}
\email{hujf@m.scnu.edu.cn}
\iqm
\sfim
\author {Bing-Nan Lu}
\email{bnlv@gscaep.ac.cn}
\gscas
\author {Jin-Yi Pang}
\email{jypang@usst.edu.cn}
\shanghai
\author {Qian Wang}
\email{qianwang@m.scnu.edu.cn, corresponding author}
\iqm
\sfim
\scnt
\OU
 
\date{\today}

\begin{abstract}
Finite-volume extrapolation is an important step for extracting physical observables from lattice calculations. However, it is a significant challenge for systems with long-range interactions. We employ symbolic regression to regress the finite-volume extrapolation formula for both short-range and long-range interactions. The regressed formula still holds the exponential form with a factor $L^n$ in front of it. The power decreases with the decreasing range of the force.  When the range of the force becomes sufficiently small, the power converges to $-1$, recovering the short-range formula as expected. Our work represents a significant advancement in leveraging machine learning to probe uncharted territories within particle physics.
\end{abstract}
\maketitle


{\it1. Introduction.}~~
Lattice quantum chromodynamics, which discretizes space into a four-dimensional hypercubic lattice and performs the calculations in a given finite volume, is a foundational computational framework for studying strong interactions. It predicts not only the spectrum of conventional hadrons but also reveals the existence of exotic hadronic states.  In particular, it has recently provided precise calculations for near-threshold states, taking the double charm tetraquark as an example.~\cite{Padmanath:2022cvl,Chen:2022vpo,Lyu:2023xro}
To extract physical observables from lattice calculations, 
 finite-volume extrapolation is essential. However, the presence of long-range interactions presents significant challenges, as they invalidate key assumptions underlying standard approaches, most notably the L\"{u}scher formula.~\cite{Luscher:1986pf}  When light particles mediate forces between scattering states, conventional methods may fail, leading to uncontrolled systematic uncertainties. Addressing these issues is crucial for reliable lattice analyses, particularly in systems where long-range effects play a dominant role, such as one-pion exchange in nucleon-nucleon scattering and the effects of light-meson exchange in the $D^*D$ system.~\cite{Meng:2023bmz} 
Recent theoretical studies~\cite{Bubna:2024izx,Dawid:2024dgy,Romero-Lopez:2020rdq,Iritani:2018vfn,Yu:2025gzg} try to solve this problem by either modifying the L\"uscher formula numerically, but without an explicit formula as simple as that for the short-range interaction,~\cite{Luscher:1985dn} or performing quantization conditions using a plane-wave basis.~\cite{Meng:2021uhz} 
 Moreover, understanding finite-volume effects in two-body systems with long-range interactions is a necessary step towards addressing similar challenges in three-body systems.

Fortunately, deriving fundamental mathematical expressions directly from empirical data~\cite{science.1165893} is an advantage of 
symbolic regression,~\cite{GeneticProgramming} which has exhibited its capability in physics and scientific research.~\cite{amil2009statistical} Unlike traditional regression, it automatically discovers interpretable laws by exploring vast expression spaces. This has enabled the rediscovery of known physical laws and the formulation of novel equations for complex systems,~\cite{Cranmer:2020wew} even when their underlying principles are elusive. 
Recent advances in symbolic regression enhance its effectiveness by incorporating physical constraints like dimensional consistency, thus improving search efficiency and model plausibility. For example, the AI Feynman algorithm has derived 100 equations from the Feynman Lectures on Physics.~\cite{Udrescu:2019mnk} Its ability to generate human-readable models accelerates discovery across various fields.~\cite{tenachi2023deep,Dong:2022trn} For instance, in high-energy physics, it has revealed fundamental distributions, such as the Tsallis distribution in hadron transverse momentum analysis.~\cite{Makke:2024whm} 
Moreover, it has  provided analytical expressions for key low-energy observables, including the Higgs mass, muon anomalous magnetic moment, and dark matter relic density.~\cite{AbdusSalam:2024obf} Reference~\cite{Lu:2022joy} uses neural network to reproduce the numerical L\"uscher’s formula to a high precision.

\begin{figure}[htbp]
    \centering
    \includegraphics[width=0.96\linewidth]{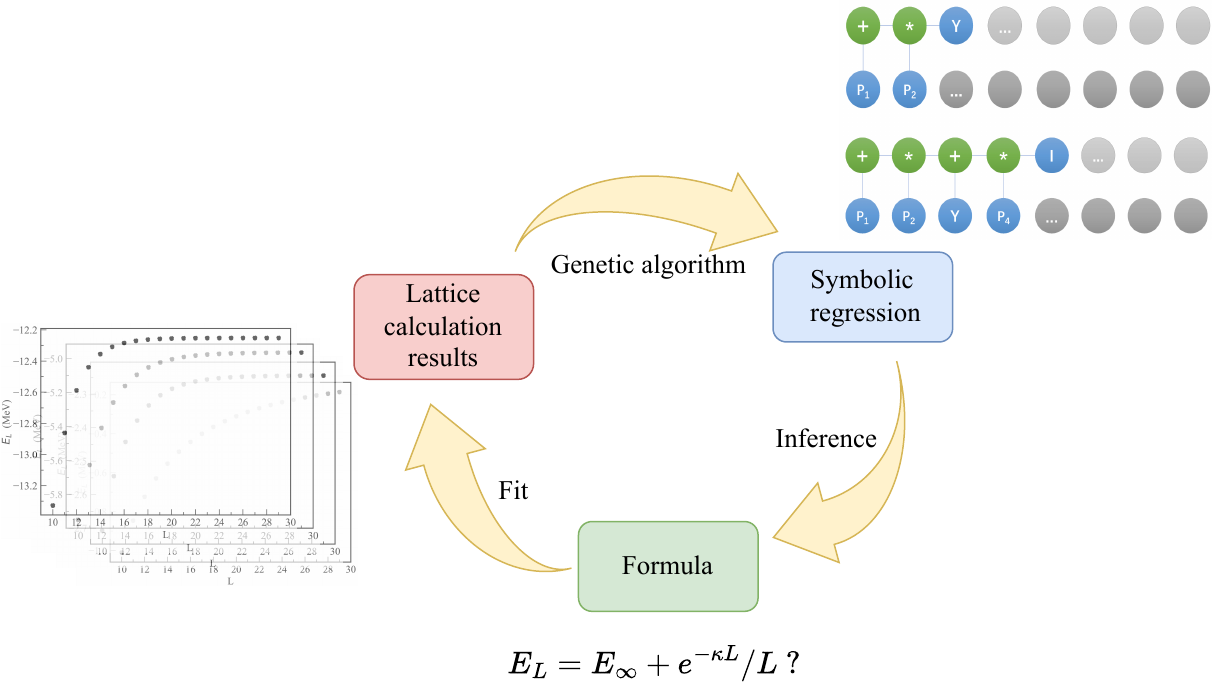}
    \caption{Workflow of using symbolic regression, LEFT, and formula.}
    \label{fig.workflow}
\end{figure}

As a result, we employ symbolic regression to probe the finite-volume extrapolation formula for long-range interactions. Before symbolic regression, we need to prepare samples for training symbolic regression. 
For these samples, we borrow the  hadron lattice effective field theory (HLEFT),~\cite{Zhang:2024yfj} which is analogous to 
the nuclear lattice effective field 
(NLEFT), a powerful computational framework for \textit{ab initio} nuclear structure calculations.~\cite{Lee:2008fa} The approach implements a systematic discretization of nucleon-nucleon interactions on a three-dimensional cubic lattice with periodic boundary conditions. These quantum many-body systems are subsequently solved using advanced numerical techniques including the Lanczos diagonalization method and auxiliary field Monte Carlo simulations. NLEFT has demonstrated remarkable success across diverse domains of nuclear physics, particularly in characterizing nuclear ground states~\cite{Borasoy:2006qn,Epelbaum:2009pd,Epelbaum:2010xt,Lahde:2013uqa,Lu:2018bat,Lu:2021tab,Elhatisari:2022zrb} and excited states,~\cite{Epelbaum:2013paa,Shen:2022bak,Meissner:2023cvo,Shen:2024qzi} elucidating nuclear intrinsic density distributions and clustering phenomena,~\cite{Epelbaum:2011md,Epelbaum:2012qn,Epelbaum:2012iu,Elhatisari:2017eno,Zhang:2024wfd} analyzing nucleus-nucleus scattering processes,~\cite{Bour:2012hn,Elhatisari:2015iga} and investigating both zero-temperature and finite-temperature nuclear matter properties.~\cite{Elhatisari:2016owd,Lu:2019nbg,Ren:2023ued,Ma:2023ahg} The framework has also been effectively extended to study hypernuclear systems.~\cite{Bour:2014bxa,Scarduelli:2020xae,Hildenbrand:2024ypw,Zhang:2024yfj} However, the presence of finite-volume effects remains a significant challenge in extending NLEFT applications to critical nuclear systems such as the neutron-rich $^6\mathrm{He}$ isotope, where the delicate balance of nuclear binding requires exceptional precision. 
This is another motivation of this work. According to the statement above, our workflow is presented in Fig.~\ref{fig.workflow}.

{\it2. Samples Generation.}~~ 
The HLEFT is used to generate the model-independent samples of two identical boson systems. These samples are fed to symbolic regression (SR), which is based on genetic algorithms (GA).~\cite{Ireland:2004kp,Fernandez-Ramirez:2008ixe} GA allows the computer to require fewer prior assumptions and provides a more flexible search space to find final expressions. Similar to the evolutionary processes in nature, GA employs mechanisms such as mutations and crossovers to generate new solutions. 
In this work, we use the famous PySR, which is a multi-population evolutionary algorithm with multiple evolutionary processes performed asynchronously. It uses a classic evolutionary algorithm for each cycle and tournament selection for individual selection. The details of the setup in PySR can be found in the Supplementary Materials (SM). 

The challenge of finite volume extrapolation lies in long-range interactions,~\cite{Bubna:2024izx,Meng:2021uhz,Dawid:2024dgy,Romero-Lopez:2020rdq,Iritani:2018vfn} especially when the 
range of force is comparable to
the box size. Luckily, the extrapolation formula for the short-range potential can be obtained from L\"{u}scher's formula for $\kappa L\gg 1$,~\cite{Luscher:1985dn} with $\kappa$ the binding momentum and $L$ the box size. 
Firstly, we regress the extrapolation formula for the short-range potential as a benchmark for the applicability  of  the PySR model. 
We work in a non-relativistic framework and consider two identical spinless particles with the same mass $m=1969~\mathrm{MeV}$. ~\footnote{The polarization of non-zero spin particles would modify the finite-volume extrapolation formula, which will not be discussed in this manuscript.}
 The short-range potential in coordinate space reads as 
\begin{equation}
    V(\boldsymbol{r})=-C_0\delta^3(\boldsymbol{r}),
    \label{Eq:short_V}
\end{equation}
where $\boldsymbol{r}$ is the relative distance between the two particles, $C_0$ is the strength of the potential for tuning the binding energy of the two-particle system. The minus sign means an attractive potential. From the effective field theory point of view, the short-range potential $V(\boldsymbol{r})$ needs a physical cutoff to isolate the long-range contribution from the short-range one. Usually, 
a single-particle regulator,~\cite{Lu:2023jyz}  $f(\boldsymbol{p}^{(\prime)}_1,\boldsymbol{p}^{(\prime)}_2)=\prod
\limits_{i=1}^2g_{\Lambda}(\boldsymbol{p_i})g_{\Lambda}(\boldsymbol{p^{\prime}_i})$ , is applied, where $g_{\Lambda}(\boldsymbol{p})=$exp$(-\boldsymbol{p}^6/2\Lambda^{6})$ is a soft cutoff function with $\boldsymbol{p_i}$ and $\boldsymbol{p^{\prime}_i}$ the momenta of the incoming and outgoing of the $i$th particle.~\cite{Zhang:2024yfj}
Thus, the Hamiltonian can be written as
\begin{equation}
H=\sum_{i=1}^2\frac{\boldsymbol{p_i}^2}{2m_i}+f(\boldsymbol{p}^{(\prime)}_1,\boldsymbol{p}^{(\prime)}_2)V(\boldsymbol{q}),
\label{Eq:H}
\end{equation}
in which $m_i$ is the mass of the $i$th particle, 
$\bm{q}=\bm{p}-\bm{p}'$ is the transferred momentum of the two particles and $\boldsymbol{p}$($\boldsymbol{p^{\prime}}$) represents relative incoming (outgoing) momentum between $\boldsymbol{p}_1$ ($\boldsymbol{p}^{\prime}_1$) and $\boldsymbol{p}_2$ ($\boldsymbol{p}^{\prime}_2$). The corresponding potential in momentum space $V(\boldsymbol{q})$ = $-C_0$ can be obtained by a Fourier transform from the expression in coordinate space.

The Schr\"{o}dinger equation is written as:
\begin{equation}
    H_L\psi=E_L\psi,
    \label{Eq:Schrodinger}
\end{equation}
where $H_L$ and $E_L$ are the Hamiltonian and the binding energy in a cubic box $L^3$. The solution of  Eq.~\eqref{Eq:Schrodinger}  can be written explicitly in a single-particle basis $|\psi\rangle=|n_1,\cdots,n_N\rangle$,~\cite{Lu:2019nbg} where $n_i$ is an integer triplet specifying the lattice coordinates. 
The matrix exact diagonalization scheme uses the implicitly restarted Lanczos method to find the eigenvalues and eigenvectors~\cite{lehoucq1998} by SciPy~\cite{Virtanen:2019joe} in Python. 
In order to study the finite volume extrapolation formula for short-range potential, we perform simulations on box sizes $L^3=10^3,11^3,\cdots,30^3~\mathrm{fm}^3$ cubic lattice with two identical particles.  The potential strength $C_0$ is set to 1.5, 2.0, 2.5, and 3.0 $\mathrm{MeV}^{-2}$ to
generate the samples. 
With this setup, we can extract the 
energy $E_L$ in finite volume  by solving Eq.~\eqref{Eq:Schrodinger} for various cases. The details of the samples can be found in the SM.

The Yukawa potential,~\cite{Bubna:2024izx}
 \begin{equation}
    V(\boldsymbol{r})=-C_{01}\delta^3(\boldsymbol{r})-C_{02} \frac{e^{-\mu r}}{r}
    \label{Eq:long_V}
\end{equation}
with $1/\mu$ reflecting the range of the force,  
is used to describe the long-range potential. 
Here, $C_{01}$ and $C_{02}$ are the strengths of the short-range potential ($\delta$-potential) and the long-range potential  (Yukawa potential). One notices that the short-range behavior of the Yukawa potential is divergent and needs regularization, which is the reason why the first term exists. For simplicity, we set $C_{01}$ = $C_{02}$ and choose the values 0.03, 0.09, 0.12 and 0.21 MeV$^{-2}$ to generate the samples.
These values allow the binding energy of the two-body system with the long-range potential to be close to that with only the short-range potential. Only in this case can one isolate the role of the long-range potential on the finite-volume extrapolation formula. For the short-range term, the single-particle regulator $f(\boldsymbol{p}^{(\prime)}_1,\boldsymbol{p}^{(\prime)}_2)$ is still applied. At the same time, for the long-range potential term, the regulator\footnote{ Here, $\Lambda$ is the cutoff to regularize the long-range potential, which is set equal to that for the short-range one above.} $\hat{f}(\boldsymbol{q})$ = exp$[-(\boldsymbol{q}^2+\mu^2)/\Lambda^2]$ is applied.~\cite{Meng:2023bmz} The Hamiltonian $H$ in this case can be written as
\begin{equation}
H=\sum_{i=1}^2\frac{\boldsymbol{p_i}^2}{2m_i}+f(\boldsymbol{p}^{(\prime)}_1,\boldsymbol{p}^{(\prime)}_2)V_S(\boldsymbol{q})+\hat{f}(\boldsymbol{q})V_L(\boldsymbol{q}).
\label{Eq:long_H}
\end{equation}
Here, both the short-range potential $V_S(\boldsymbol{q})$ = $-C_{01}$ and the long-range potential $V_L(\boldsymbol{q})\sim\frac{1}{\boldsymbol{q}^2+\mu^2}$ can be transformed into their coordinate forms by Fourier transformation. 
We perform simulations on box sizes $L^3=10^3,11^3,\cdots,30^3$ fm$^3$ cubic lattices. 
 We set the potential strength 
$C_{01}=C_{02}=0.03,~0.09,~0.12, \text{and}~0.21 ~\mathrm{MeV}^{-2}$ 
to obtain binding energies similar to those for only the short-range potential. The details can be found in the SM.

 
 {\it3. Symbolic Regression.}~~We perform symbolic regression on the generated samples that characterize finite-volume effects. The PySR model~\cite{Cranmer:2023pysr}  samples the space of analytic expressions defined by the set of operators, input variables, and constant terms for minimization through genetic programming. The pool of operations includes  addition, subtraction, multiplication, division, exponential, logarithm, square, etc. The input variable is box size $L$. Unlike the massive data used in deep learning, a few samples should be sufficient in the PySR model. The model evolution process employs several mechanisms, such as mutations and crossovers, to generate new expressions. The algorithm performs mutations over 50 iterations of 50 different population samples, with each population containing 35 individuals. There are two elements to measure the goodness of the output formula in the PySR model, i.e., loss and score. The value of loss is used to measure how well the output formula describes the samples, which is defined as mean square error,
\begin{equation}
\mathrm{Loss}=\sum_{i=1}^{N} (E_{\mathrm{PySR}}(L_{i}) - E_L({L_{i}}))^2/N.
\label{Eq:loss}
\end{equation}
Here, $E_{\mathrm{PySR}}(L)$ is the output formula of PySR model, $E_L({L})$ is the energy at box size $L^3$ calculated from lattice effective field theory (LEFT). 
Score is used to estimate the form of the formula, which rewards minimal loss and penalizes the more complicated formula. It can be defined as
\begin{equation}
\mathrm{Score}=-\frac{\Delta\ \mathrm{ln(Loss)} }{\Delta\ C}.
\label{Eq:score}
\end{equation}
Here, $C$ is the complexity~\footnote{The upper limit of the $C$ value can be set larger to allow higher order contributions as that in analytic analysis ~\cite{Luscher:1985dn}.} (its values can be found in the SM), which is defined as the total number of operations, variables, and constants used in a formula. The PySR model will consider both score, i.e. complexity and loss,  to choose the best formula. The other  parameters used in  training process can be found in the SM. 

The training process of symbolic regression is very similar to biological population inheritance. In initialization,  the PySR model randomly provides several kinds of evolutionary frameworks, which are different from each other and are used as a starting point for a given evolution. Mutations and crossovers occur throughout each iteration. During each  iteration, formulae will be produced and the PySR model will choose the best formula for the next iteration according to the value of loss and score. The iterations will stop until the loss and complexity reach the specified values set at the beginning. During this process, the PySR model will choose the best formula across all iterations.
\begin{figure}[htbp]
\centering
\includegraphics[width=0.48\textwidth]{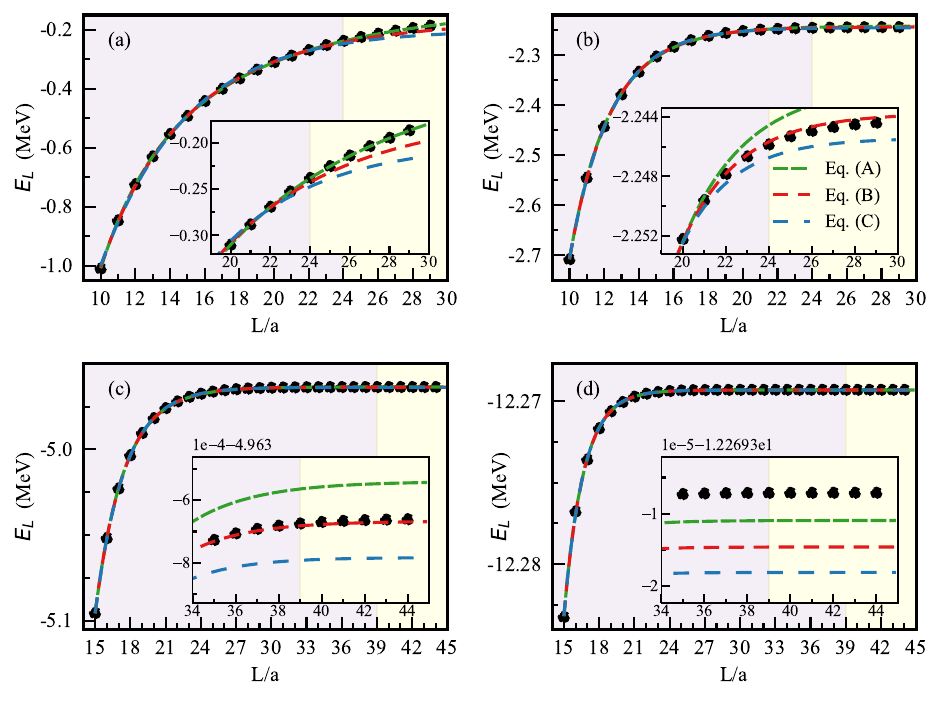}
    \caption{The results of fitting the formula to samples. There are four cases of (a) $C_0=1.5$ MeV$^{-2}$, (b) $C_0=2.0$ MeV$^{-2}$, (c) $C_0=2.5$ MeV$^{-2}$ and (d) $C_0$ = 3.0 MeV$^{-2}$. Equations (a), (b) and (c) correspond to Eqs.~\eqref{Eq:A_short}, \eqref{Eq:B_short} and \eqref{Eq:C_short}, respectively. The box size $L$ ranges from 10 to 30 fm. The purple region and the yellow region indicate the fitted and predicted areas by Eqs. (a), (b) and (c), respectively. Each graph has its own sub-graph to show greater detail of the lines between the fitted and predicted areas.}
    \label{fig:short_FVE_Eb}
\end{figure}

{\it4. Results.}~~Based on the PySR model, we can perform a symbolic regression for the samples of the short-range attractive potential  to find an expression that uniformly describes the finite-volume extrapolation.  
 In the four cases of $C_0$ values, three formulae are accepted by the PySR model,~\footnote{For the case of $C_0=1.5~\mathrm{MeV}^{-2}$, PySR model gives the formula $E_L=C'_1+C'_2/L+C'_3/L^2$. For the purpose of uniformity of the formula, when the ratio $C^\prime_2/C_3^\prime$ is small enough, the formula is approximated as $E_L=C_1+C_2e^{-C_3L}/L^2$.} as follows:
\begin{subequations}
\begin{align}
    E_L &= C_1+C_2e^{-C_3L}/L^2, \tag{8a}\label{Eq:A_short} \\
    E_L &= C_1+C_2e^{-C_3L}/L,\tag{8b}\label{Eq:B_short} \\
    E_L &= C_1+C_2e^{-C_3L}. \tag{8c}\label{Eq:C_short}
\end{align}
\end{subequations}
In Fig.~\ref{fig:short_FVE_Eb}, we show the results of the three formulas fitting to the samples with the short-range potential, generated from LEFT.  The purple area is the fitting region for the formulae derived from the PySR model. The yellow area is the predicted one. For the shallow bound state, e.g., Fig.~\ref{fig:short_FVE_Eb}(a), equation~\eqref{Eq:A_short} is the best choice. However, the volume size is not big enough to see the convergent behavior, making the regressed formula not universal. When the binding energy increases to a few MeV, i.e., Figs.~\ref{fig:short_FVE_Eb}(b) and ~\ref{fig:short_FVE_Eb}(c), equation~\eqref{Eq:B_short} is better than Eq.~\eqref{Eq:A_short}. When the binding energy further increases to more than $10~\mathrm{MeV}$,  e.g., Fig.~\ref{fig:short_FVE_Eb}(d), 
all three formulae are good 
 choices because their loss values are much smaller than those of the other sub-figures of Fig.~\ref{fig:short_FVE_Eb}.  As a result, we conclude that Eq.~\eqref{Eq:B_short} is the best regressed uniform  formula to describe the samples with a few MeV binding energy.

The current well-established theoretical works~\cite{Luscher:1986pf,Doring:2018xxx,Hammer:2017uqm,Konig:2017krd} are based on the L\"uscher formula~\cite{Luscher:1985dn} for the short-range potential with $\kappa L\gg 1$, where $\kappa$ is the binding momentum of the two-body system. The energy shift of a shallow two-body bound state of identical bosons in finite volume is 
 \begin{equation}
E_L=E_{\infty}+\frac{C^\prime}{L}\exp{(-\kappa L)},
\label{Eq:Luscher_formula}
\end{equation}
where $C^\prime$ is the free parameter.
This formula is exactly the form of Eq.~\eqref{Eq:B_short} regressed by PySR. In addition, 
the fitted values of the parameter $C_3$ in Eq.~\eqref{Eq:B_short} for the last three cases of Fig.~\ref{fig:short_FVE_Eb} are equal to the binding momentum $\kappa$ numerically. This also demonstrates that we have successfully regressed the finite-volume extrapolation formula for the short-range potential. 
\begin{figure}[htbp]
\centering
    \includegraphics[width=0.48\textwidth]{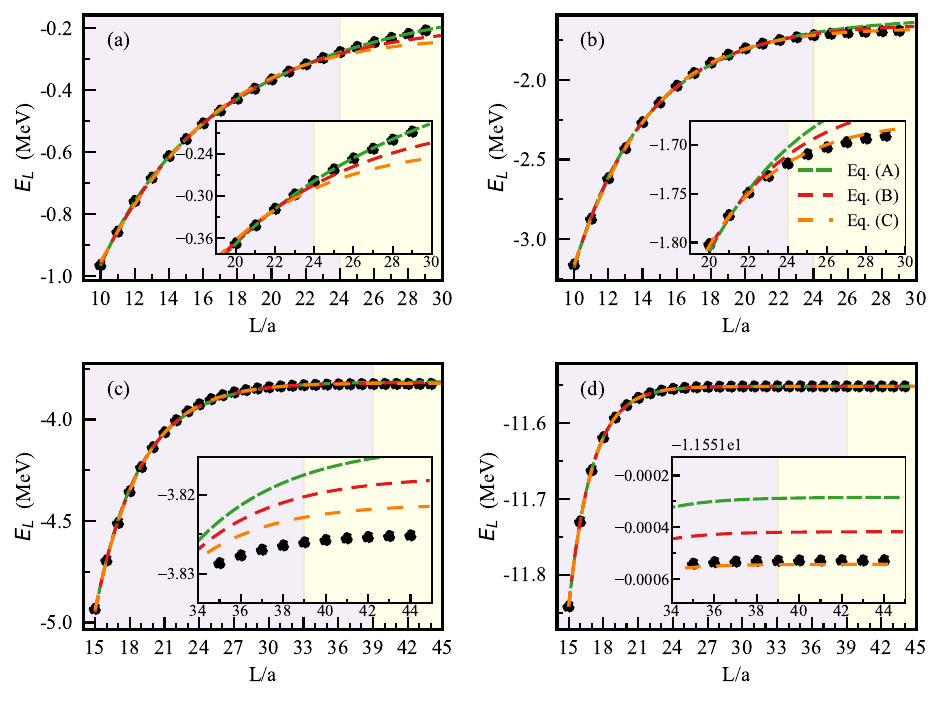}
    \caption{The results of  Eq.~\eqref{Eq:A_long} (green short dashed curves), Eq.~\eqref{Eq:B_long} (red dashed curves), and Eq.~\eqref{Eq:C_long} (orange long dashed curves) fitting to the samples of the long-range potentials with (a) $C_{01}=C_{02}=0.03~\mathrm{MeV}^{-2}$ , (b) $C_{01}=C_{02}=0.09~\mathrm{MeV}^{-2}$ , (c) $C_{01}=C_{02}=0.12~\mathrm{MeV}^{-2}$ , and (d)  $C_{01}=C_{02}=0.21~\mathrm{MeV}^{-2}$ , respectively. The box size $L$ ranges from 10 to 30$~\mathrm{fm}$. The purple region and the yellow region indicate the fitted and predicted areas. Each graph has its own subgraph to show more details around the boundary between fitted and predicted areas. The black points are the samples generated from LEFT.}
    \label{fig:long_FVE_Eb}
\end{figure}

For the long-range potential,
the strengths of the attractive potentials are set to obtain similar binding energies for the short-range potential. The range of the force parameter $\mu$ is set as  $\mu=20~\mathrm{MeV}$ for these four $C_0$ cases. The other values of the force parameter and their effect on the finite-volume extrapolation 
will be discussed afterwards.
The following three formulae~\footnote{For the case of $C_{01}=C_{02}=$ 0.03 MeV$^{-2}$, PySR model gets the formula $E_L=C'_1+C'_2/L+C'_3L$. For the purpose of a uniform form of the formula, when the ratio $C^\prime_3/C^\prime_2$ is small enough, the formula is approximated as $E_L=C_1+C_2e^{-C_3L}/L$.}
\begin{subequations}
\begin{align}
    E_L &= C_1+C_2 e^{-C_3 L}/L,\tag{10a}\label{Eq:A_long} \\
    E_L &= C_1+C_2e^{-C_3L},\tag{10b}\label{Eq:B_long} \\
    E_L &= C_1+C_2e^{-C_3L}L. \tag{10c}\label{Eq:C_long}
\end{align}
\end{subequations}
are accepted by PySR model. Figure~\ref{fig:long_FVE_Eb} shows the fitting results of the above three formulae to the energy with the long-range interaction. For the case with binding energy smaller than $1~\mathrm{MeV}$, i.e., figure~\ref{fig:long_FVE_Eb}(a), equation~\eqref{Eq:A_long} is preferred. However, the binding energy is too small to make the energy convergent in box size $30^3~\mathrm{fm}^3$, making Eq.~\eqref{Eq:A_long} inapplicable to other cases. 
For the other cases, equation~\eqref{Eq:C_long} describes the samples very well and can be considered as a uniform formula to describe the energy shift with the range of the potential  as large as $10~\mathrm{fm}$.  In comparison with the formula for the short-range potential, the power of $L$ in front of the exponential becomes larger. This enlightens us that the power of $L$ has a correlation with the range of the force.  
 A theoretical formulation of the two-body quantization condition in the presence of a long-range force was proposed in Refs.~\cite{Bubna:2024izx,Meng:2021uhz,Dawid:2024dgy,Romero-Lopez:2020rdq,Iritani:2018vfn,Yu:2025gzg}. We expect that a comparable power-law dependence on $L$ can be obtained within the improved L\"{u}scher formula.
 
As discussed above, the power of $L$ in front of the exponential is related to the range of the force. Accordingly, we study the tendency of this power in the following. According to the results from the machine learning approach, the formula can be written 
 as a general form,  
\begin{equation}
    E_L = C_1+C_2e^{-C_3L}L^n.
    \label{Eq:long_range}
\end{equation}
Here, $n$ is an unknown parameter and needs to be fixed for various samples. 
The parameter $\mu$ in Eq.~\eqref{Eq:long_V} is set as $10,~20,~\dots,120,~400,~600,~800~\mathrm{MeV}$ corresponding to the range of force $19.73,~9.87,\dots, ~1.64,~0.49,~0.33,~0.25~\mathrm{fm}$. The first two ranges of force are compatible with the box size.
As analyzed in the SM, 
 the power of $L$ in Eq.~\eqref{Eq:long_range} decreases when the force range decreases. 
Moreover, when $\mu$ goes to infinity, i.e., the short-range interaction, the power $n$ comes back to the formula 
$E_L = C_1+C_2\exp(-C_3L)L^{-1}$ deduced from the L\"{u}scher formula.~\cite{Luscher:1985dn} It indicates that no matter the form of the potential, the finite-volume extrapolation formula $E_L = C_1+C_2\exp(-C_3L)L^{-1}$ can be reproduced once the range of force is short enough. 
On the other hand, when $\mu$ goes to zero, the power goes to infinity, which presents the behavior for the infinite long-range interaction (see the SM). The range of the force for the first two points of Fig.S2 of SM is compatible with the box size and cannot reflect the correct finite-volume behavior. 


{\it5. Summary and Outlook.}
~~Finite-volume extrapolation is an important step for extracting physical observables from lattice calculations. 
However, it is a significant challenge for systems with long-range interactions, especially when the range of the force is comparable to the lattice box size. Several theoretical works have been put forward to extract the finite-volume extrapolation formula based on the modified L\"uscher formula numerically. 
To obtain an exact extrapolation formula, we employ symbolic regression for both short-range and long-range interactions. 
The samples are generated by the HLEFT and are fed to symbolic regression. 
As a benchmark, the finite-volume extrapolation formula for short-range interaction has been successfully regressed, i.e.,  the exponential  form $\exp{(-\kappa L)}$ with the $L^{-1}$ factor in front of it, 
which is in good agreement with the theoretical result. 
Furthermore, we turn to the case of long-range interactions. The regressed results still keep the exponential form, but with the power of $L$ dependent on the range of the force, i.e.,  the power of $L$ decreases with the decreasing range of the force. When the range of the force becomes sufficiently small, the power converges to $-1$, recovering the short-range formula as expected. 
This agreement further validates the correctness of our extrapolation formula for long-range forces. 
 Our work represents a significant advancement in leveraging machine learning to probe uncharted territories within particle physics.

{\it Acknowledgments.}~~
We thank Serdar Elhatisari, Feng-Kun Guo, Ulf-G. Mei{\ss}ner, Andreas Nogga, and Akaki G. Rusetsky for their useful discussions on
the finite volume dependence from the theoretical side. 
This work was supported in part 
by the  National Natural Science Foundation of China  (Grant Nos.~12375072,~12375073,~12275259, and~12135011).
The work of Q.W. was also supported by
Guangdong Provincial Funding (Grant No.~2019QN01X172). B.N. Lu is also supported by  the National Security Academic Fund  (Grant No.U2330401).

\nocite{*}

\begin{thebibliography}{bstutf8}
\bibitem{Padmanath:2022cvl} Padmanath M and Prelovsek S 2022 Phys. Rev. Lett. 129 032002
\bibitem{Chen:2022vpo} Chen S, Shi C, Chen Y, Gong M, Liu Z, Sun W and Zhang R 2022 Phys. Lett. B 833 137391
\bibitem{Lyu:2023xro} Lyu Y, Aoki S, Doi T, Hatsuda T, Ikeda Y and Meng J 2023 Phys. Rev. Lett. 131 161901
\bibitem{Luscher:1986pf} L\"uscher M 1986 Commun. Math. Phys. 105 153–188
\bibitem{Meng:2023bmz} Meng L, Baru V, Epelbaum E, Filin A A and Gasparyan A M 2024 Phys. Rev. D 109 L071506
\bibitem{Bubna:2024izx} Bubna R, Hammer H-W, Müller F, Pang J-Y, Rusetsky A and Wu J-J 2024 JHEP 05 168
\bibitem{Dawid:2024dgy} Dawid S M, Romero-López F and Sharpe S R 2025 JHEP 01 060
\bibitem{Romero-Lopez:2020rdq} Romero-López F, Rusetsky A, Schlage N and Urbach C 2021 JHEP 02 060
\bibitem{Iritani:2018vfn} Iritani T, Aoki S, Doi T, Hatsuda T, Ikeda Y, Inoue T, Ishii N, Nemura H and Sasaki K 2019 JHEP 03 007
\bibitem{Yu:2025gzg} Yu K, Wang G-J, Wu J-J and Yang Z 2025 JHEP 04 108
\bibitem{Luscher:1985dn} Lüscher M 1986 Commun. Math. Phys. 104 177
\bibitem{Meng:2021uhz} Meng L and Epelbaum E 2021 JHEP 10 051
\bibitem{science.1165893} Schmidt M and Lipson H 2009 Science 324 81–85
\bibitem{GeneticProgramming} Koza J R 1992 Genetic programming: on the programming of computers by means of natural selection (MIT Press USA)
\bibitem{amil2009statistical} Amil N M, Bredeche N, Gagné C, Gelly S, Schoenauer M and Teytaud O 2009 Genetic Programming (Springer US) p. 327–338
\bibitem{Cranmer:2020wew} Cranmer M, Sanchez-Gonzalez A, Battaglia P, Xu R, Cranmer K, Spergel D and Ho S 2020 arXiv:2006.11287 [cs.LG]
\bibitem{Udrescu:2019mnk} Udrescu S-M and Tegmark M 2020 Sci. Adv. 6 eaay2631
\bibitem{tenachi2023deep} Tenachi W, Ibata R and Diakogiannis F I 2023 Astrophys. J. 959 99
\bibitem{Dong:2022trn} Dong Z, Kong K, Matchev K T and Matcheva K 2023 Phys. Rev. D 107 055018
\bibitem{Makke:2024whm} Makke N and Chawla S 2024 PNAS Nexus 3 467
\bibitem{AbdusSalam:2024obf} AbdusSalam S, Abel S and Crispim Romão M 2025 Phys. Rev. D 111 015022
\bibitem{Lu:2022joy} Lu Y, Wang Y-J, Chen Y and Wu J-J 2024 Chin. Phys. C 48 073101
\bibitem{Zhang:2024yfj} Zhang Z, Hu X-Y, He G, Liu J, Shi J-A, Lu B-N and Wang Q 2025 Phys. Rev. D 111 036002
\bibitem{Lee:2008fa} Lee D 2009 Prog. Part. Nucl. Phys. 63 117–154
\bibitem{Borasoy:2006qn} Borasoy B, Epelbaum E, Krebs H, Lee D and Meißner U-G 2007 Eur. Phys. J. A 31 105–123
\bibitem{Epelbaum:2009pd} Epelbaum E, Krebs H, Lee D and Meißner U-G 2010 Phys. Rev. Lett. 104 142501
\bibitem{Epelbaum:2010xt} Epelbaum E, Krebs H, Lee D and Meißner U-G 2010 Eur. Phys. J. A 45 335–352
\bibitem{Lahde:2013uqa} L\"ahde T A, Epelbaum E, Krebs H, Lee D, Meißner U-G and Rupak G 2014 Phys. Lett. B 732 110–115
\bibitem{Lu:2018bat} Lu B-N, Li N, Elhatisari S, Lee D, Epelbaum E and Meißner U-G 2019 Phys. Lett. B 797 134863
\bibitem{Lu:2021tab} Lu B-N, Li N, Elhatisari S, Ma Y-Z, Lee D and Meißner U-G 2022 Phys. Rev. Lett. 128 242501
\bibitem{Elhatisari:2022zrb} Elhatisari S and others 2024 Nature 630 59–63
\bibitem{Epelbaum:2013paa} Epelbaum E, Krebs H, Lähde T A, Lee D, Meißner U-G and Rupak G 2014 Phys. Rev. Lett. 112 102501
\bibitem{Shen:2022bak} Shen S, Elhatisari S, L\"ahde T A, Lee D, Lu B-N and Meißner U-G 2023 Nature Commun. 14 2777
\bibitem{Meissner:2023cvo} Meißner U-G, Shen S, Elhatisari S and Lee D 2024 Phys. Rev. Lett. 132 062501
\bibitem{Shen:2024qzi} Shen S, Elhatisari S, Lee D, Meißner U-G and Ren Z 2024 Phys. ReV. Lett. 134 162503
\bibitem{Epelbaum:2011md} Epelbaum E, Krebs H, Lee D and Meißner U-G 2011 Phys. Rev. Lett. 106 192501
\bibitem{Epelbaum:2012qn} Epelbaum E, Krebs H, L\"ahde T A, Lee D and Meißner U-G 2012 Phys. Rev. Lett. 109 252501
\bibitem{Epelbaum:2012iu} Epelbaum E, Krebs H, L\"ahde T A, Lee D and Meißner U-G 2013 Phys. Rev. Lett. 110 112502
\bibitem{Elhatisari:2017eno} Elhatisari S, Epelbaum E, Krebs H, L\"ahde T A, Lee D, Li N, Lu B-N, Meißner U-G and Rupak G 2017 Phys. Rev. Lett. 119 222505
\bibitem{Zhang:2024wfd} Zhang S, Elhatisari S, Meißner U-G and Shen S 2024 arXiv:2411.17462 [nucl-th]
\bibitem{Bour:2012hn} Bour S, Hammer H-W, Lee D and Meißner U-G 2012 Phys. Rev. C 86 034003
\bibitem{Elhatisari:2015iga} Elhatisari S, Lee D, Rupak G, Epelbaum E, Krebs H, L\"ahde T A, Luu T and Meißner U-G 2015 Nature 528 111
\bibitem{Elhatisari:2016owd} Elhatisari S and others 2016 Phys. Rev. Lett. 117 132501
\bibitem{Lu:2019nbg} Lu B-N, Li N, Elhatisari S, Lee D, Drut J E, L\"ahde T A, Epelbaum E and Meißner U-G 2020 Phys. Rev. Lett. 125 192502
\bibitem{Ren:2023ued} Ren Z, Elhatisari S, L\"ahde T A, Lee D and Meißner U-G 2024 Phys. Lett. B 850 138463
\bibitem{Ma:2023ahg} Ma Y-Z, Lin Z, Lu B-N, Elhatisari S, Lee D, Li N, Meißner U-G, Steiner A W and Wang Q 2024 Phys. Rev. Lett. 132 232502
\bibitem{Bour:2014bxa} Bour S, Lee D, Hammer H-W and Meißner U-G 2015 Phys. Rev. Lett. 115 185301
\bibitem{Scarduelli:2020xae} Scarduelli V, Gasques L R, Chamon L C and Lépine-Szily A 2020 Eur. Phys. J. A 56 24
\bibitem{Hildenbrand:2024ypw} Hildenbrand F, Elhatisari S, Ren Z and Meißner U-G 2024 Eur. Phys. J. A 60 215
\bibitem{Ireland:2004kp} Ireland D G, Janssen S and Ryckebusch J 2004 Nucl. Phys. A 740 147–167
\bibitem{Fernandez-Ramirez:2008ixe} Fernández-Ramírez C, Moya de Guerrra E, Udías A and Udías J M 2008 Phys. Rev. C 77 065212
\bibitem{Lu:2023jyz} Lu B-N and Deng B-G 2023 arXiv:2308.14559 [nucl-th]
\bibitem{lehoucq1998} Lehoucq R B, Sorensen D C and Yang C 1998 ARPACK users’ guide: solution of large-scale eigenvalue problems with implicitly restarted Arnoldi methods (SIAM USA)
\bibitem{Virtanen:2019joe} Virtanen P and others 2020 Nature Meth. 17 261
\bibitem{Cranmer:2023pysr} Cranmer M 2023 arXiv:2305.01582 [astro-ph.IM]
\bibitem{Doring:2018xxx} D\"oring M, Hammer H-W, Mai M, Pang J-Y, Rusetsky A. and Wu J 2018 Phys. Rev. D 97 114508
\bibitem{Hammer:2017uqm} Hammer H-W, Pang J-Y and Rusetsky A 2017 JHEP 09 109
\bibitem{Konig:2017krd} K\"onig S and Lee D 2018 Phys. Lett. B 779 9–15


\end{thebibliography}

\onecolumngrid
\newpage
\appendix

\clearpage


\end{document}


\include{def-com}
\title{\boldmath Supplementary Material}

\author {Wei-Jie Zhang}
\iqm
\moe
\sfim
\author {Zhenyu Zhang} \email{Co-first author}
\iqm
\moe
\sfim
\author {Jifeng Hu}
\email{hujf@m.scnu.edu.cn}
\iqm
\sfim
\author {Bing-Nan Lu}
\email{bnlv@gscaep.ac.cn}
\gscas
\author {Jin-Yi Pang}
\email{jypang@usst.edu.cn}
\shanghai
\author {Qian Wang}
\email{qianwang@m.scnu.edu.cn, corresponding author}
\iqm
\sfim
\scnt
\OU

\date{\today}

\maketitle

In the supplementary material, we present the details of the PySR model, the intermediate results during the training and some theoretical discussions.

\section{The details of the PySR model}\label{Appendix:SR_para}

Lots of parameters, which can be modified to meet requirements, are provided in PySR. Some vital parameters will be introduced below. After one training process, PySR gives one formula as output. During this period, a new equation accompanied by its complexity and loss, is generated in each iteration.

Here, complexity is used to estimate the form of the formula. The more complicated equation, the higher value of complexity. The formula is consisted by constant term, variable term and sign of operators. Thus PySR provides some parameters to modify them. Parameter ``complexity\_of\_constants'' is used to modify complexity of the constants in formula and parameter ``complexity\_of\_variables'' is used to modify the variables, namely $L$ in the formula. In the training process,``complexity\_of\_variables'' is changed larger than its default value. While ``complexity\_of\_constants'' remain default. Sign of operators is used to separate the operators into  two kinds of operators. If the operators, such as $``+"$ and $``\times"$, need two elements, it is called binary operator, denoted by  ``binary\_operators''. Besides, if the operators, such as ``exp'' and ``log'', need only one element, it is called unary operator, denoted as ``unary\_operators''. After  carefully considering final results, it is not appropriate to use trigonometric functions to characterize finite volume effect. Thus,  parameter ``unary\_operators'' = [``exp'', ``sqrt'', ``log''] is used. To simplify the result, some complicated forms, such as nested structure,  should be prohibited by the parameter ``nested\_constraints''. Here, we set parameter ``nested\_constraints'' = ``exp'':\{``exp'':0\}, which means the form of nested exponential functions is not allowed.

Except for the complexity, the value of loss also plays an important role in choosing formula. Like most machine learning processes, it is necessary to use loss function to judge which formula is better. Parameter ``elementwise\_loss'' is used to select the loss function in the training period. Usually, mean square error (MSE) is used in machine learning. Thus, ``elementwise\_loss'' = L2DistLoss().

After defining complexity and loss, PySR will select a formula considering two elements mentioned above. Parameter ``model\_selection'' is used to control which formula can be chosen by PySR. Here, we set the ``model\_selection'' = ``best'', which means PySR will consider both complexity and loss to choose the best formula.
\begin{table}[htbp]
\caption{Some parameters and their functions and values in PySR model.}
    \centering
    \renewcommand{\arraystretch}{1.5}
    \begin{tabular}{c p{10cm}<{\centering} c}
    \hline
    \hline
    \textbf{Parameter}&\multicolumn{1}{c}{\textbf{Function}}&\textbf{Value}  \\
    \hline
    populations&determine the number of populations in each generation during evolution&50\\
    population\_size&number of the candidate in each population &35\\
    ncycles\_per\_iteration&number of total mutations to run, per 10 samples of the population, per iteration&550\\
    niterations&number of iterations of the algorithm to run&40\\
    maxsize&upper complexity limit of final result&15\\
    binary\_operators&list of strings for binary operators used in the search&``$+$'', ``$-$'', ``$*$'', ``$/$''\\
    unary\_operators&operators which only take a single scalar as input& ``exp'', ``sqrt'', ``log''\\
    nested\_constraints&specifies how many times a combination of operators can be nested&``exp'':{``exp'':0}\\
    complexity\_of\_variables&global complexity of variables&1\\
    complexity\_of\_constants&complexity of constants&2\\
    optimizer\_iterations&number of iterations that the constants optimizer can take&30\\
   perturbation\_factor&constants are perturbed by a max factor of (perturbation\_factor*T + 1), either multiplied by this or divided by this&0.076\\
   parsimony&multiplicative factor for how much to punish complexity&0.032\\
   adaptive\_parsimony\_scaling&weigh of simple formula&20.0\\
    model\_selection&select a final expression from the list of best expression at each complexity&``best''\\
    elementwise\_loss&elementwise loss funciton&``L2DistLoss()''\\ 
    \hline
    \hline
    \end{tabular}
    \label{tab:SR_para}
\end{table}

\section{The samples from lattice EFT}
\label{app:details_on_lattice}

In two-body system, considering only short-range potential between two identical particles, we use the Lattice Effective Field Theory (LEFT) to generate the $E_L$ at finite volume for different cases. The detailed data can be found Tab.~\ref{tab:short_range}. The lattice spacing $a$ should not be set smaller than the typical hadron size, as hadrons are degrees of freedom in LEFT. The spatial lattice spacing is chosen as $a=1/200(1/300)~\mathrm{MeV}^{-1}\approx 0.99(0.67)~\mathrm{fm}$ for smaller (larger) binding energies. 
The soft cutoff is chosen $\Lambda=350 (600)~\mathrm{MeV}$ to make sure that it is much larger than the binding momentum $\kappa$ and smaller than $\pi/a$. To satisfy the requirement of the lattice EFT discussed above, we choose $a=1/200~\mathrm{MeV}^{-1}$ and $\Lambda=350~\mathrm{MeV}$ for 
$C_0=1.5,~2.0~\mathrm{MeV}^{-2}$. 
Those values for $C_0=2.5,~3.0~\mathrm{MeV}^{-2}$ are 
$a=1/300~\mathrm{MeV}^{-1}$ and $\Lambda=600~\mathrm{MeV}$. Those setups require minimal computing resources and makes the physics independent on them. 

\begin{table}[htbp]
 \caption{The energy $E_L$ for box size $L^3$ in different spatial lattice spacings $a$, cutoff $\Lambda$, $C_0$, which adjusts the strength of the $\delta$-potential,  and $L/a$. The masses of particles are set as 1969 MeV.}
  \centering
  \begin{tabular}{p{1.5cm}<{\centering}p{1.5cm}<{\centering}p{1.5cm}<{\centering}p{1.5cm}<{\centering}p{1.5cm}<{\centering}p{1cm}<{\centering}p{1.5cm}<{\centering}p{1.5cm}<{\centering}p{1.5cm}<{\centering}p{1.5cm}<{\centering}}
    \hline
    \hline
    {$a$(MeV$^{-1}$)}  &1/200&{1/200}&{1/300}&{1/300}&&1/200&{1/200}&{1/300}&{1/300}\\
    {$\Lambda$(MeV)}   &350  &{350}  &{600}  &{600}  &&350  &{350}  &{600}  &{600}  \\
    {$C_0$(MeV$^{-2}$)}&1.50 &{2.00} &{2.50} &{3.00} &&1.50 &{2.00} &{2.50} &{3.00} \\ 
   \hline
   $L/a$&\multicolumn{4}{c}{$E_L$(MeV)}&$L/a$&\multicolumn{4}{c}{$E_L$(MeV)}\\
    10&-1.010&-2.708&\multicolumn{2}{c}{$\cdots$} &28&-0.194&-2.245               &-4.964&-12.269\\
    11&-0.846&-2.545&\multicolumn{2}{c}{$\cdots$} &29&-0.186&-2.244               &-4.964&-12.269\\
    12&-0.724&-2.443&\multicolumn{2}{c}{$\cdots$} &30&\multicolumn{2}{c}{$\cdots$}&-4.964&-12.269\\
    13&-0.629&-2.379&\multicolumn{2}{c}{$\cdots$} &31&\multicolumn{2}{c}{$\cdots$}&-4.964&-12.269\\
    14&-0.553&-2.334&\multicolumn{2}{c}{$\cdots$} &32&\multicolumn{2}{c}{$\cdots$}&-4.964&-12.269\\
    15&-0.492&-2.303               &-5.096&-12.283&33&\multicolumn{2}{c}{$\cdots$}&-4.964&-12.269\\
    16&-0.442&-2.284               &-5.052&-12.277&34&\multicolumn{2}{c}{$\cdots$}&-4.964&-12.269\\
    17&-0.400&-2.271               &-5.023&-12.274&35&\multicolumn{2}{c}{$\cdots$}&-4.964&-12.269\\
    18&-0.365&-2.262               &-5.003&-12.272&36&\multicolumn{2}{c}{$\cdots$}&-4.964&-12.269\\
    19&-0.335&-2.256               &-4.990&-12.271&37&\multicolumn{2}{c}{$\cdots$}&-4.964&-12.269\\
    20&-0.310&-2.252               &-4.982&-12.270&38&\multicolumn{2}{c}{$\cdots$}&-4.964&-12.269\\
    21&-0.288&-2.250               &-4.976&-12.270&39&\multicolumn{2}{c}{$\cdots$}&-4.964&-12.269\\
    22&-0.269&-2.248               &-4.972&-12.270&40&\multicolumn{2}{c}{$\cdots$}&-4.964&-12.269\\
    23&-0.252&-2.247               &-4.969&-12.269&41&\multicolumn{2}{c}{$\cdots$}&-4.964&-12.269\\
    24&-0.238&-2.245               &-4.968&-12.269&42&\multicolumn{2}{c}{$\cdots$}&-4.964&-12.269\\
    25&-0.225&-2.246               &-4.966&-12.269&43&\multicolumn{2}{c}{$\cdots$}&-4.964&-12.269\\
    26&-0.214&-2.245               &-4.966&-12.269&44&\multicolumn{2}{c}{$\cdots$}&-4.964&-12.269\\
    27&-0.203&-2.245               &-4.965&-12.269\\
    \hline
    \hline
  \end{tabular}
  \label{tab:short_range}
\end{table}
\begin{table}[htbp]
 \caption{The energy $E_L$ in box size $L^3$ with different spatial lattice spacings $a$, cutoff $\Lambda$, $C_{01}$, which adjusts the size of short-range potential, $C_{02}$, which adjusts the size of long-range potential, and $L/a$. The masses of the two particles are set as 1969 MeV.}
  \centering
    \begin{tabular}{p{2cm}<{\centering}p{1.5cm}<{\centering}p{1.5cm}<{\centering}p{1.5cm}<{\centering}p{1.5cm}<{\centering}p{1cm}<{\centering}p{1.5cm}<{\centering}p{1.5cm}<{\centering}p{1.5cm}<{\centering}p{1.5cm}<{\centering}}
    \hline
    \hline
    {$a$(MeV$^{-1}$)}            &{1/200}&{1/200}&{1/300}&{1/300}& &{1/200}&{1/200}&{1/300}&{1/300}\\
    {$\Lambda$(MeV)}             &{350}  &{350}  &{600}  &{600}  & &{350}  &{350}  &{600}  &{600}  \\
    {$C_{01}/C_{02}$(MeV$^{-2}$)}&{0.03} &{0.09} &{0.12} &{0.21} & &{0.03} &{0.09} &{0.12} &{0.21} \\ 
    \hline
    $L/a$&\multicolumn{4}{c}{$E_L$(MeV)}&$L/a$&\multicolumn{4}{c}{$E_L$(MeV)}\\
    10&-0.965&-3.162&\multicolumn{2}{c}{$\cdots$} &28&-0.220&-1.694               &-3.856&-11.552\\
    11&-0.857&-2.873&\multicolumn{2}{c}{$\cdots$} &29&-0.208&-1.691               &-3.848&-11.552\\
    12&-0.758&-2.619&\multicolumn{2}{c}{$\cdots$} &30&\multicolumn{2}{c}{$\cdots$}&-3.842&-11.552\\
    13&-0.683&-2.430&\multicolumn{2}{c}{$\cdots$} &31&\multicolumn{2}{c}{$\cdots$}&-3.838&-11.552\\
    14&-0.614&-2.266&\multicolumn{2}{c}{$\cdots$} &32&\multicolumn{2}{c}{$\cdots$}&-3.834&-11.552\\
    15&-0.559&-2.142               &-4.933&-11.842&33&\multicolumn{2}{c}{$\cdots$}&-3.832&-11.552\\
    16&-0.509&-2.037               &-4.694&-11.730&34&\multicolumn{2}{c}{$\cdots$}&-3.830&-11.552\\
    17&-0.467&-1.957               &-4.510&-11.662&35&\multicolumn{2}{c}{$\cdots$}&-3.829&-11.552\\
    18&-0.429&-1.891               &-4.354&-11.619&36&\multicolumn{2}{c}{$\cdots$}&-3.828&-11.552\\
    19&-0.397&-1.841               &-4.234&-11.593&37&\multicolumn{2}{c}{$\cdots$}&-3.827&-11.552\\
    20&-0.367&-1.802               &-4.137&-11.576&38&\multicolumn{2}{c}{$\cdots$}&-3.826&-11.552\\
    21&-0.342&-1.772               &-4.063&-11.566&39&\multicolumn{2}{c}{$\cdots$}&-3.826&-11.552\\
    22&-0.318&-1.749               &-4.004&-11.561&40&\multicolumn{2}{c}{$\cdots$}&-3.826&-11.552\\
    23&-0.298&-1.732               &-3.960&-11.557&41&\multicolumn{2}{c}{$\cdots$}&-3.826&-11.552\\
    24&-0.279&-1.719               &-3.926&-11.555&42&\multicolumn{2}{c}{$\cdots$}&-3.825&-11.552\\
    25&-0.262&-1.710               &-3.900&-11.554&43&\multicolumn{2}{c}{$\cdots$}&-3.825&-11.552\\
    26&-0.247&-1.703               &-3.881&-11.553&44&\multicolumn{2}{c}{$\cdots$}&-3.825&-11.552\\
    27&-0.233&-1.697               &-3.867&-11.552\\
    \hline
    \hline
    \end{tabular}
  \label{tab:long_range}
\end{table}

In two-body system, considering both the short-range and the long-range potential between two particles, the LEFT is used to generate the energy $E_L$ in $L^3$ box for different cases. The range of the force parameter $\mu$ in is set as $\mu= 20~\mathrm{MeV}$. 
 The spatial lattice spacing is chosen as $a=1/200 (1/300)~\mathrm{MeV}^{-1}$ for $C_{01}=C_{02}=0.03,~0.09 ~\mathrm{MeV}^{-2} (0.12,~0.21 ~\mathrm{MeV}^{-2})$ corresponding to $\Lambda=350~\mathrm{MeV}    (600~\mathrm{MeV})$. 
 More detailed samples can be found in Tab.~\ref{tab:long_range}.

To study the correlation between the power of $L$ and the range of the force, we generate samples with different ranges of the force. 
To remove the effect of the binding energy on the value of $n$, the parameters $C_{01}$ and $C_{02}$ are chosen to produce binding energy around $-3.83~\mathrm{MeV}$. 
The other parameters are set as 
$a=1/300~\mathrm{MeV}^{-1}$, $\Lambda=600~\mathrm{MeV}$ and box size $L$ ranges from $10\sim 30~\mathrm{fm}$ (i.e. $L/a$ ranges from 15 to 45). The parameter $\mu$ is set as $10,~20,~\dots,120,~400,~600,~800~\mathrm{MeV}$ corresponding the range of force $19.73,~9.87,\dots, ~1.64,~0.49,~0.33,~0.25~\mathrm{fm}$. The first two ranges of force are compatible with the box size.
More specific, 
in two-body system, considering both short-range and the long-range potential, we generate samples with 15 $\mu$ values. The detailed results can be found in Tab.~\ref{tab:short_long_range}.
\begin{table}[htbp]
\caption{The energy $E_L$ in $L^3$ box size with $a=1/300~\mathrm{MeV}^{-1}$ and $\Lambda=600~\mathrm{MeV}$.
The long-range parameter $\mu$ is set in the range $[10,~120]~\mathrm{MeV}$. }
    \centering
    \renewcommand{\arraystretch}{1.5}
    
    \begin{tabular}{p{2.5cm}<{\centering}|p{1.1cm}<{\centering}p{1.1cm}<{\centering}p{1.1cm}<{\centering}p{1.1cm}<{\centering}p{1.1cm}<{\centering}p{1.1cm}<{\centering}p{1.1cm}<{\centering}p{1.1cm}<{\centering}p{1.1cm}<{\centering}p{1.1cm}<{\centering}p{1.1cm}<{\centering}p{1.1cm}<{\centering}}
    \hline     
    \hline
        $\mu$(MeV)&10&20&30&40&50&60&70&80&90&100&110&120\\
        $C_{01}/C_{02}$(MeV$^{-2}$)&0.107&0.120&0.133&0.147&0.161&0.175&0.191&0.206&0.223&0.240&0.258&0.276\\
        \hline
         $L/a$&\multicolumn{12}{c}{$E_L$(MeV)}\\
         15&-5.041&-4.933&-4.819&-4.749&-4.667&-4.571&-4.554&-4.470&-4.454&-4.414&-4.390&-4.341\\
         16&-4.800&-4.694&-4.587&-4.527&-4.456&-4.371&-4.366&-4.291&-4.285&-4.253&-4.237&-4.194\\
         17&-4.610&-4.510&-4.410&-4.361&-4.300&-4.224&-4.229&-4.162&-4.164&-4.138&-4.128&-4.090\\
         18&-4.446&-4.354&-4.266&-4.228&-4.178&-4.111&-4.125&-4.065&-4.074&-4.053&-4.048&-4.014\\
         19&-4.317&-4.235&-4.157&-4.130&-4.088&-4.029&-4.050&-3.995&-4.009&-3.993&-3.992&-3.960\\
         20&-4.208&-4.137&-4.070&-4.053&-4.020&-3.967&-3.995&-3.944&-3.962&-3.950&-3.951&-3.921\\
         21&-4.124&-4.063&-4.006&-3.997&-3.971&-3.923&-3.955&-3.907&-3.929&-3.919&-3.922&-3.894\\
         22&-4.054&-4.004&-3.956&-3.955&-3.934&-3.890&-3.926&-3.880&-3.905&-3.896&-3.902&-3.874\\
         23&-4.000&-3.960&-3.919&-3.924&-3.907&-3.866&-3.909&-3.861&-3.888&-3.881&-3.887&-3.860\\
         24&-3.957&-3.926&-3.891&-3.901&-3.888&-3.849&-3.891&-3.848&-3.876&-3.869&-3.877&-3.850\\
         25&-3.924&-3.901&-3.870&-3.885&-3.874&-3.836&-3.880&-3.838&-3.867&-3.861&-3.869&-3.843\\
         26&-3.898&-3.881&-3.855&-3.872&-3.864&-3.828&-3.873&-3.831&-3.861&-3.856&-3.864&-3.838\\
         27&-3.879&-3.867&-3.844&-3.864&-3.856&-3.821&-3.867&-3.826&-3.856&-3.852&-3.860&-3.835\\
         28&-3.863&-3.856&-3.836&-3.857&-3.851&-3.816&-3.863&-3.823&-3.853&-3.849&-3.858&-3.832\\
         29&-3.852&-3.848&-3.830&-3.852&-3.847&-3.813&-3.861&-3.820&-3.851&-3.847&-3.856&-3.830\\
         30&-3.843&-3.842&-3.825&-3.849&-3.844&-3.811&-3.859&-3.818&-3.849&-3.845&-3.854&-3.829\\
         31&-3.837&-3.838&-3.822&-3.847&-3.842&-3.809&-3.857&-3.817&-3.848&-3.844&-3.853&-3.828\\
         32&-3.832&-3.834&-3.820&-3.845&-3.841&-3.808&-3.856&-3.816&-3.847&-3.843&-3.853&-3.827\\
         33&-3.828&-3.832&-3.818&-3.843&-3.840&-3.807&-3.855&-3.815&-3.847&-3.843&-3.852&-3.827\\
         34&-3.825&-3.830&-3.817&-3.842&-3.839&-3.807&-3.855&-3.815&-3.846&-3.842&-3.852&-3.826\\
         35&-3.823&-3.829&-3.816&-3.842&-3.838&-3.806&-3.854&-3.814&-3.846&-3.842&-3.851&-3.826\\
         36&-3.822&-3.828&-3.815&-3.841&-3.838&-3.805&-3.854&-3.814&-3.846&-3.842&-3.851&-3.826\\
         37&-3.821&-3.827&-3.814&-3.841&-3.838&-3.805&-3.854&-3.814&-3.846&-3.842&-3.851&-3.826\\
         38&-3.820&-3.827&-3.814&-3.841&-3.837&-3.805&-3.854&-3.814&-3.846&-3.842&-3.851&-3.826\\
         39&-3.819&-3.826&-3.814&-3.841&-3.837&-3.805&-3.854&-3.814&-3.845&-3.841&-3.851&-3.826\\
         40&-3.818&-3.826&-3.814&-3.840&-3.837&-3.805&-3.854&-3.814&-3.845&-3.841&-3.851&-3.826\\
         41&-3.818&-3.826&-3.813&-3.840&-3.837&-3.805&-3.853&-3.814&-3.845&-3.841&-3.851&-3.826\\
         42&-3.818&-3.825&-3.813&-3.840&-3.837&-3.804&-3.853&-3.814&-3.845&-3.841&-3.851&-3.826\\
         43&-3.818&-3.825&-3.813&-3.840&-3.837&-3.804&-3.853&-3.813&-3.845&-3.841&-3.851&-3.826\\
         44&-3.817&-3.825&-3.813&-3.840&-3.837&-3.804&-3.853&-3.813&-3.845&-3.841&-3.851&-3.826\\
    \hline
    \hline
    \end{tabular}
    \label{tab:short_long_range}
\end{table}

\section{The details of Symbolic regression}

The details of the output formula for each iteration in the PySR model for the short-range potential and the long-range potential are listed in  Tab.~\ref{tab:short_B} and Tab.~\ref{tab:long_B}, respectively.

\begin{table*}[htbp]
\caption{The process of PySR model generates formula with the samples for  $a=1/200$ MeV$^{-1}$, $\Lambda=350$ MeV and $C_0=2.0$  MeV$^{-2}$ short-range potential only. The first column represents the evolution times. The last column is the output formula for each evolution after the simplification. And the bold formula is accepted by PySR model after one cycle.}
    \centering
    \begin{tabular}{p{2cm}<{\centering} p{2cm}<{\centering} p{3cm}<{\centering} p{2cm}<{\centering}p{6cm}<{\centering}}
    \hline
    \hline
         Times&Complexity&Loss&Score&Equation  \\
         \hline
         1 &1 &$5.818$             &0.000&$L$\\
         2 &2 &$0.014$             &6.016&$-2.312$\\
         3 &4 &$0.010$             &0.180&$L-2.410$\\
         4 &5 &$0.007$             &0.396&$-2.549\exp(L)$\\
         5 &6 &$0.005$             &0.220&$-3.144\exp(\sqrt{L})$\\
         6 &7 &$0.003$             &0.653&$(-0.028/L)-1.992$\\
         7 &8 &$0.003$             &0.000&$(-0.028/L)-1.992$\\
         8 &9 &$0.001$             &0.682&$-0.001/L^2-2.148$\\
         9 &10&$3.679\times10^{-4}$&1.341&$(\sqrt{L}-0.171)/L-3.696$\\
         10&11&$1.831\times10^{-6}$&5.302&$-29.889\exp(-83.458L)-2.245$\\
         11&12&$1.826\times10^{-6}$&0.003&$-\exp(-83.388L+3.393)-2.245$\\
         12&13&$6.333\times10^{-7}$&1.059&$\bm{-0.629\exp(-66.049L)/L-2.244}$\\
         13&14&$6.292\times10^{-7}$&0.006&$-0.628\exp(-66.022L)/L-2.244$\\
         \hline
         \hline
    \end{tabular}
    \label{tab:short_B}
\end{table*}
\begin{table*}[htbp]
\caption{The process of generating a formula with the samples for $a=1/200$ MeV$^{-1}$, $\Lambda=350$ MeV and $C_{01}=C_{02}=0.09$ MeV$^{-2}$ short-range and long-range potential. The bold formula is accepted by PySR model after one cycle. The last line indicates that, even when the complexity increases, the output formula is not significantly improved.}
    \centering
    \begin{tabular}{p{2cm}<{\centering} p{2cm}<{\centering} p{3cm}<{\centering} p{2cm}<{\centering}p{6cm}<{\centering}}
    \hline
    \hline
         Times&Complexity&Loss&Score&Equation  \\
         \hline
         1&1  &$4.660$             &0.000&$L$\\
         2&2  &$0.153$             &3.418&$\log(L)$\\
         3&4  &$0.072$             &0.380&$-0.172/L$\\
         4&5  &$0.021$             &1.237&$-0.613/\sqrt{L}$\\
         5&6  &$0.008$             &1.006&$-\exp(0.058/L)$\\
         6&8  &$0.002$             &0.676&$-\exp(0.056/L)-L$\\
         7&9  &$0.002$             &0.253&$-0.004/L^2-1.413$\\
         8&11 &$5.672\times10^{-5}$&1.646&$-15.311/\exp(46.272L)-1.661$\\
         9&13 &$6.928\times10^{-6}$&1.051&$\bm{-695.626L/\exp(63.072L)-1.677}$\\
         10&15&$6.571\times10^{-6}$&0.026&$-681.508L/\exp(62.467L)+L^2-1.660$\\
         \hline
         \hline
    \end{tabular}
    \label{tab:long_B}
\end{table*}

\section{The effect of range of the force on the finite-volume extrapolation}

\begin{figure*}[htbp]
\centering
    \includegraphics[width=0.32\textwidth]{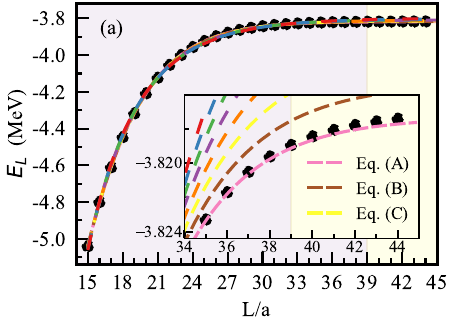}
    \includegraphics[width=0.32\textwidth]{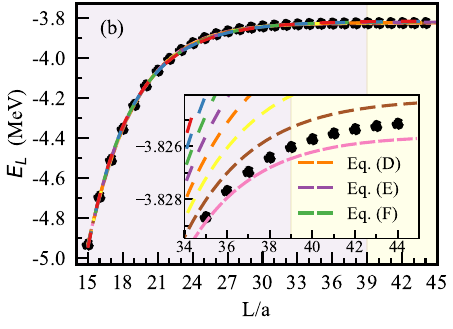}
    \includegraphics[width=0.32\textwidth]{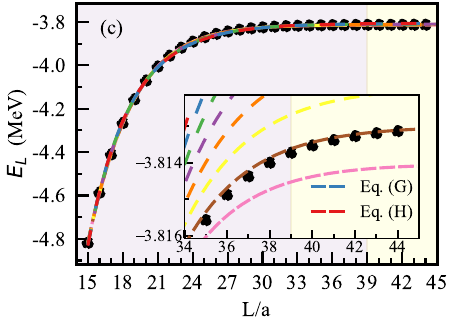}

    \includegraphics[width=0.32\textwidth]{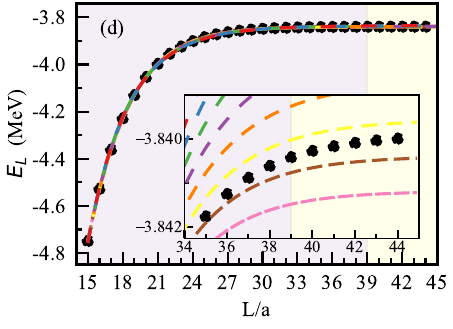}
    \includegraphics[width=0.32\textwidth]{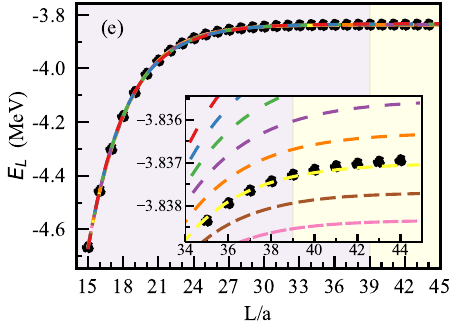}
    \includegraphics[width=0.32\textwidth]{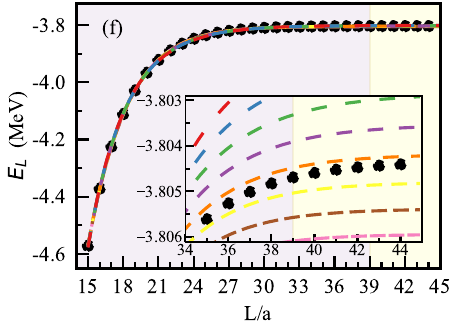}

    \includegraphics[width=0.32\textwidth]{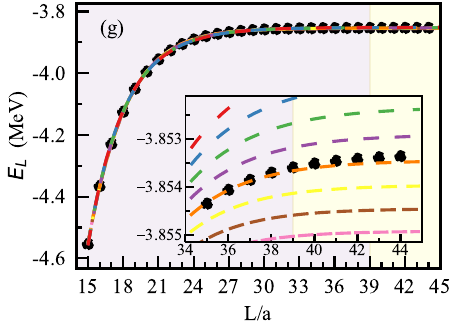}
    \includegraphics[width=0.32\textwidth]{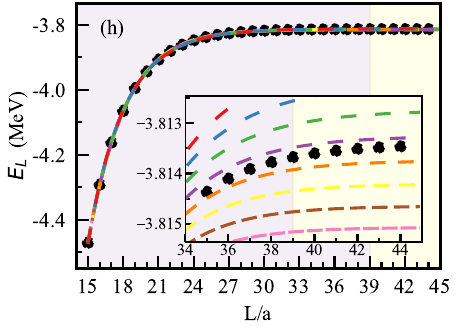}
    \includegraphics[width=0.32\textwidth]{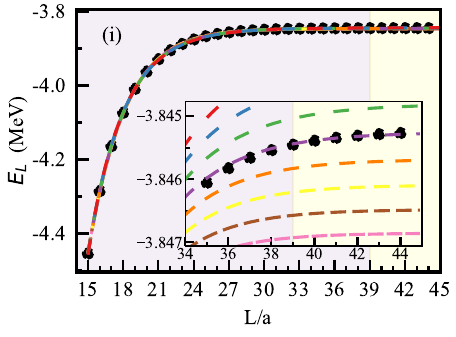}

    \includegraphics[width=0.32\textwidth]{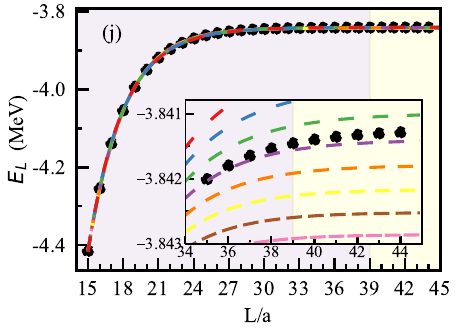}
    \includegraphics[width=0.32\textwidth]{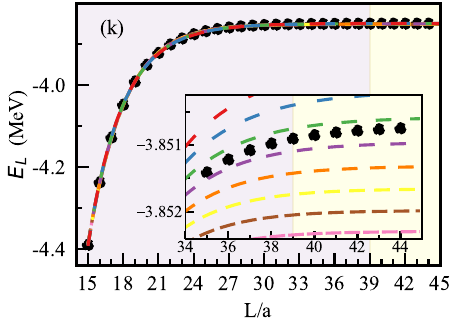}
    \includegraphics[width=0.32\textwidth]{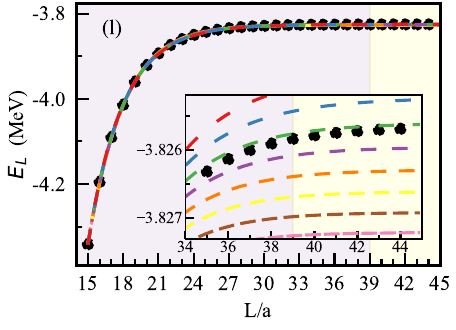}

    \includegraphics[width=0.32\textwidth]{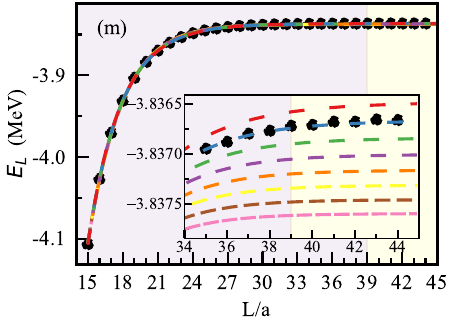}
    \includegraphics[width=0.32\textwidth]{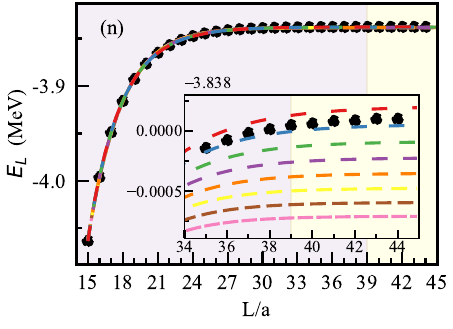}
    \includegraphics[width=0.32\textwidth]{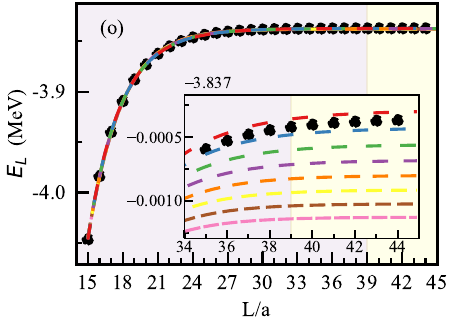}
    \caption{The fitting results of $E_L = C_1+C_2e^{-C_3L}L^n$ to samples with the long-range parameter $\mu=10,~20,\dots,~120,~400,~600,~800 \mathrm{MeV}$ denoted from (a) to (o) in order. 
    For all the cases, the binding energy is around $3.83~\mathrm{MeV}$ and the parameters $a=1/300~\mathrm{MeV}^{-1}$,  $\Lambda= 600~\mathrm{MeV}$.
    The eight lines, i.e. pink, brown, yellow, orange, purple, green, blue and red lines, represent the power $n=5/2,~2,~3/2,~1,~1/2,~0,~-1/2,~-1$ of $L$ from bottom to top in each figure. The black dots are the samples  generated by LEFT for various cases.} 
    \label{fig:long_FVE_range}
\end{figure*}

\begin{figure}[htbp]
\centering
    \includegraphics[width=0.48\textwidth]{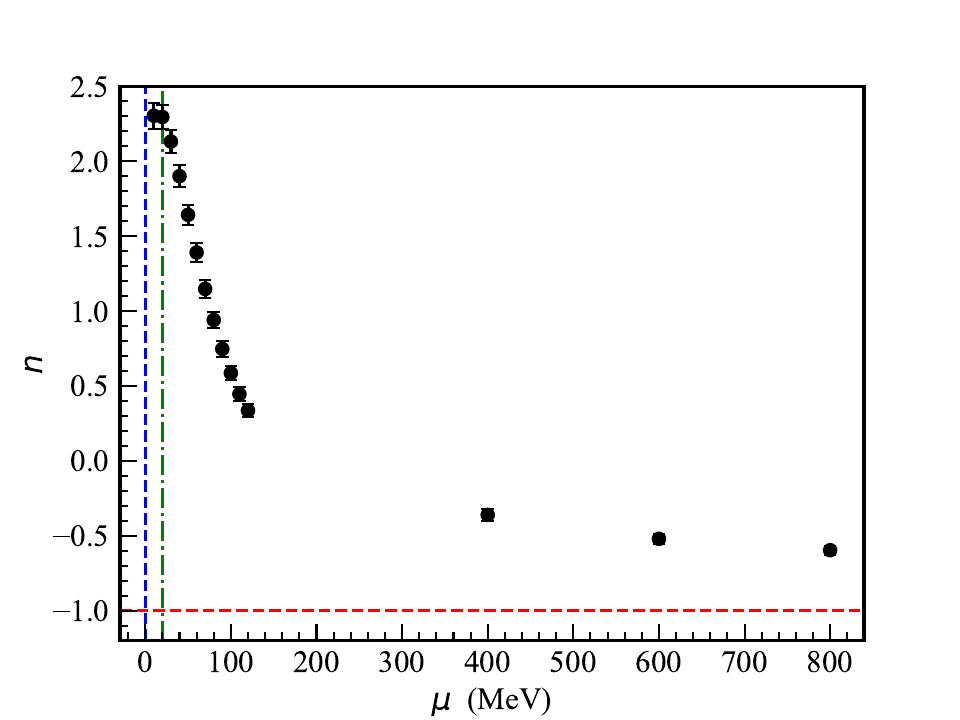}
    \caption{The dependence of the power $n$ in Eq.~\eqref{Eq:long_range} on the range parameter $\mu$. The vertical blue dashed line is the infinity long-range limit. The vertical green dot-dashed line corresponds to the lower limit of our box size, i.e. $ \frac{\hbar c}{10~\mathrm{fm}}$. The horizontal red dashed line is the short-range value $-1$ from L\"uscher formula, which is also the convergent value for infinity large $\mu$. The behavior presents very good short-range and long-range behaviors.}
    \label{fig:long_range_power_n}
\end{figure}

\begin{figure}[htbp]
\centering
    \includegraphics[width=0.48\textwidth]{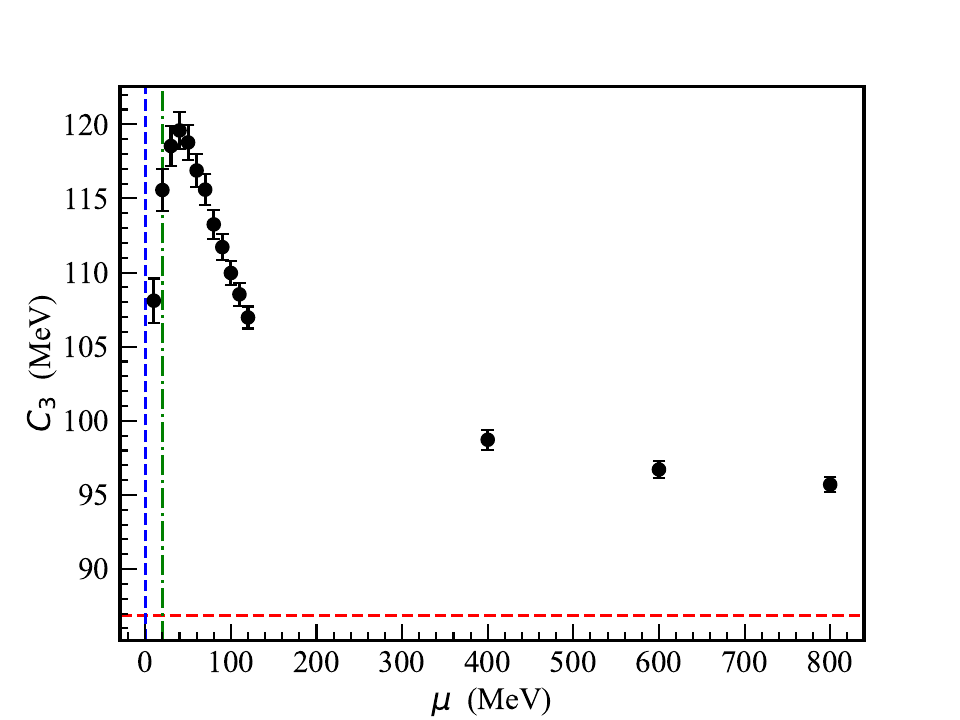}
    \caption{The dependence of the parameter $C_3$ on the range parameter $\mu$ with the power $n$ corresponding to the results of Fig.~\ref{fig:long_range_power_n}. The vertical blue dashed line is the infinity long-range limit. The vertical green dot-dashed line corresponds to the lower limit of our box size, i.e. $ \frac{\hbar c}{10~\mathrm{fm}}$. The horizontal red dashed line is the binding momentum $\kappa=\sqrt{mE_\infty}$, which is also the convergent value for infinity large $\mu$.}
    \label{fig:long_range_para_C3}
\end{figure}

As discussed in the letter, the finite-volume extrapolation formula for both short-range and long-range interactions can be written in a compact form
\begin{equation}
    E_L = C_1+C_2e^{-C_3L}L^n.
    \label{Eq:long_range}
\end{equation}
with the power of $L$ depending on the range of the force. We set the range parameter 
$\mu=10,~20,~\dots,120,~400,~600,~800$ $\mathrm{MeV}$ corresponding the range of force $19.73,~9.87,\dots, ~1.64,~0.49,~0.33,~0.25~\mathrm{fm}$ to probe the behavior of $n$. Fig.~\ref{fig:long_FVE_range} presents the fitting results for the eight $n=5/2,~2,~3/2,~1,~1/2,~0,~-1/2,~-1$ (denoting as Eq.(A), Eq.(B),$\dots$ Eq.(H) in Fig.~\ref{fig:long_FVE_range}) values to the energy shift. Here the fitting parameters are the $C_1$, $C_2$ and $C_3$ in Eq.~\eqref{Eq:long_range}. One can see that the $n$ value decreases with the increasing $\mu$. Furthermore, we release the power $n$ as an additional  free parameter in the fitting. The behavior of the power $n$ in terms of the range parameter $\mu$ is illustrated in Fig.~\ref{fig:long_range_power_n}. One can see when $\mu$ goes to infinity, i.e. the short-range limit, the power $n$ comes back to $-1$, recovering  the formula 
$E_L = C_1+C_2\exp(-C_3L)L^{-1}$ deduced from the L\"{u}scher formula for the short-range interaction.   
On the other hand, when $\mu$  goes to zero, the power goes to infinity, which presents the behavior for the infinity long-range interaction. The flat behavior of the first two points is limited by our box size. In this case, we also present the dependence of the parameter $C_3$ on the range parameter $\mu$ in Fig.~\ref{fig:long_range_para_C3}. The value of $C_3$ goes to the binding momentum for short-range interaction, i.e. for infinity $\mu$ value.   


\nocite{*}

\onecolumngrid
\newpage
\appendix